\documentstyle[12pt,worldsci]{article}
 1
\def\Re{{\cal R \mskip-4mu \lower.1ex \hbox{\it e}\,}}
\def\Im{{\cal I \mskip-5mu \lower.1ex \hbox{\it m}\,}}
\def\ie{{\it i.e.}}
\def\eg{{\it e.g.}}

\def\etal{{\it et al.}}
\def\ibid{{\it ibid}.}
\def\sub#1{_{\lower.25ex\hbox{$\scriptstyle#1$}}}
\def\sul#1{_{\kern-.1em#1}}
\def\sll#1{_{\kern-.2em#1}}
\def\sbl#1{_{\kern-.1em\lower.25ex\hbox{$\scriptstyle#1$}}}
\def\ssb#1{_{\lower.25ex\hbox{$\scriptscriptstyle#1$}}}
\def\sbb#1{_{\lower.4ex\hbox{$\scriptstyle#1$}}}

\def\gev{\,{\rm GeV}}
\def\mev{\,{\rm MeV}}
\def\to{\rightarrow}
\def\tcf{\ifmmode \tau cF \else $\tau$cF\fi}
\def\dmix{\ifmmode D^0-\bar D^0 \else $D^0-\bar D^0$\fi}
\def\dm{\ifmmode \Delta m_D \else $\Delta m_D$\fi}
\def\mh{\ifmmode m\sbl H \else $m\sbl H$\fi}
\def\mch{\ifmmode m_{H^\pm} \else $m_{H^\pm}$\fi}
\def\mt{\ifmmode m_t\else $m_t$\fi}
\def\mc{\ifmmode m_c\else $m_c$\fi}
\def\mz{\ifmmode M_Z\else $M_Z$\fi}
\def\mw{\ifmmode M_W\else $M_W$\fi}
\def\mws{\ifmmode M_W^2 \else $M_W^2$\fi}
\def\mhs{\ifmmode m_H^2 \else $m_H^2$\fi}
\def\mzs{\ifmmode M_Z^2 \else $M_Z^2$\fi}
\def\mts{\ifmmode m_t^2 \else $m_t^2$\fi}
\def\mcs{\ifmmode m_c^2 \else $m_c^2$\fi}
\def\mchs{\ifmmode m_{H^\pm}^2 \else $m_{H^\pm}^2$\fi}
\def\ztwo{\ifmmode Z_2\else $Z_2$\fi}
\def\zone{\ifmmode Z_1\else $Z_1$\fi}
\def\mtwo{\ifmmode M_2\else $M_2$\fi}
\def\mone{\ifmmode M_1\else $M_1$\fi}
\def\tb{\ifmmode \tan\beta \else $\tan\beta$\fi}
\def\xw{\ifmmode x\sub w\else $x\sub w$\fi}
\def\ch{\ifmmode H^\pm \else $H^\pm$\fi}
\def\lum{\ifmmode {\cal L}\else ${\cal L}$\fi}
\def\inpb{\ifmmode {\rm pb}^{-1}\else ${\rm pb}^{-1}$\fi}
\def\infb{\ifmmode {\rm fb}^{-1}\else ${\rm fb}^{-1}$\fi}
\def\epem{\ifmmode e^+e^-\else $e^+e^-$\fi}
\def\ppb{\ifmmode \bar pp\else $\bar pp$\fi}

\newskip\zatskip \zatskip=0pt plus0pt minus0pt
\def\matth{\mathsurround=0pt}

\def\atversim#1#2{\lower0.7ex\vbox{\baselineskip\zatskip\lineskip\zatskip
  \lineskiplimit 0pt\ialign{$\matth#1\hfil##\hfil$\crcr#2\crcr\sim\crcr}}}

\begin{document}
\rightline{\vbox{\halign{&#\hfil\cr
&SLAC-PUB-6695\cr
&October 1994\cr
&T/E\cr}}}
\begin{center}{{\bf THE CHARM PHYSICS POTENTIAL OF A TAU-CHARM
FACTORY\footnote{Work supported by
the Department of
Energy, Contract DE-AC03-76SF00515}
\footnote{To appear in {\it The Tau-Charm Factory in the Era of
B-Factories and CESR}, Stanford, CA, August 15-16, 1994}
\\}
\vglue 0.6cm
{J.L. HEWETT \\}
\baselineskip=13pt
{\it Stanford Linear Accelerator Center,
Stanford University, Stanford, CA   94309\\}
\vglue 0.2cm
{\tenrm ABSTRACT}}
\end{center}
{\rightskip=3pc
 \leftskip=3pc
 \baselineskip=12pt
 \noindent
The charm physics program accessible to a Tau-Charm Factory
is summarized.  Semileptonic,
leptonic, hadronic, rare and forbidden charm decay modes are discussed,
as well as \dmix\ mixing and CP violation in the charm sector.  The
theoretical expectations in the Standard Model and beyond as well as the
experimental capabilities of a Tau-Charm Factory
are examined for each of the above processes.
\vglue 0.6cm}
\section{Overview}

One of the outstanding problems in particle physics
is the origin of the fermion mass and mixing spectrum.  At present the
best approach in addressing this question is to study the properties of all
heavy fermions in detail.  While investigations of the $K$ and $B$ systems
have and will continue to play a central role in the quest to understand
flavor physics, in-depth examinations of the charm-quark sector have yet to
be performed, leaving a gap in our knowledge.  Since charm is the only heavy
charged $+2/3$ quark presently accessible to experiment, it provides the
sole window of opportunity to examine flavor physics in this sector.  In
addition, charm allows a complimentary probe of Standard Model (SM) physics
(and beyond) to that attainable from the down-quark sector.

Detailed measurements of heavy quark systems are best realized at
high precision, high luminosity machines.  Here we discuss the charm
physics potential at a high luminosity (${\cal L}=10^{33}\ {\rm cm}^{-2}\
{\rm sec}^{-1}$) \epem\ collider operating at the $c\bar c$ and $\tau\bar
\tau$ threshold region, \ie, a Tau-Charm Factory (\tcf ).  Such a machine
would facilitate an in-depth analysis of the charm-quark sector
(as well as $\tau$-lepton\cite{perl} and charmonium\cite{toki} studies)
without the contamination of $b$-quark production.
The high luminosity is absolutely essential in order to accomplish the physics
goals of the machine.  In some cases it is necessary in order to
measure a reaction at the high level of precision that is desired, while in
other cases, high luminosity is required to reach the very small transition
rates.

The exclusive nature of charm production at a \tcf\ automatically supplies
the production kinematics and low combinatoric backgrounds that
are essential to background rejection.  The charm production cross
section in this energy region is large and well measured, as can be seen
in Table 1 from Ref.\ [3].  Although the largest production rates
occur at threshold, the proposed $B$-Factories and a high luminosity $Z$
Factory could produce a comparable (but slightly smaller) number of
charmed particles.  The advantage of the threshold region at a \tcf\
is that it provides
the capability to control backgrounds and systematic errors.  Since the beam
energy can be tuned to lie just below or above production threshold, physics
backgrounds can be directly measured, instead of estimated via Monte
Carlo simulations.  Because heavy flavors are pair produced at threshold,
the observation of the decay of one particle, cleanly tags its partner.
This yields a sample\cite{rafe} of $10^6 - 10^7$ singly tagged $D$-mesons
per year, a size which is crucial in order to obtain highly precise
measurements of branching fractions and to search for rare processes.
\begin{table}
\centering
\begin{tabular}{|c|c|c|c|} \hline\hline
$\sqrt s$ GeV & Particle & Cross Section (nb) & Pairs Produced $(\times 10^8)$
\\ \hline
3.77 & $D^0\bar D^0$ & $5.8\pm 0.8$ & 1.0 \\
3.77 & $D^+D^-$ & $4.2\pm 0.7$ & 0.8 \\
4.03 & $D_s\bar D_s$ & $0.7\pm 0.2$ & 0.24 \\
4.14 & $D_s\bar D^*_s$ & $0.9\pm 0.2$ & 0.32 \\ \hline\hline
\end{tabular}
\caption{Charm production cross sections and rates.}
\end{table}

The charm physics program at a \tcf\ is strong and diverse.
Charm hadrons possess a rich variety of weak decays: Cabibbo allowed,
single Cabibbo suppressed, double Cabibbo suppressed, leptonic, semileptonic,
and rare higher-order decays.  The scale of the interactions is in the middle
of the regime where perturbative and non-perturbative QCD effects are cleanly
separable.  Hence the charm system  provides an excellent laboratory to
test our understanding of QCD, the dynamics of heavy quark decay, and
the structure of the hadronic weak current.
The knowledge gained from a comprehensive study
would supply the needed benchmarks for lattice QCD and result in a more
reliable extrapolation to the $B$ system.  Fundamental parameters in
the SM, such as the values of the Cabbibo-Kobayashi-Maskawa (CKM) mixing
matrix elements involving charm-quark
decay, would also be more precisely determined.  The charm system also
offers the opportunity to study CP violation in the up-quark sector and to
search for new sources of CP violation.
Loop induced processes are very sensitive to new
particles which may participate in the loop and hence may provide a signature
for physics beyond the SM.  Rare charm interactions compliment searches in the
down-quark sector because the
couplings to new particles may be flavor dependent, either through mass
dependent couplings or through mixing angles.

We now turn to a discussion of semileptonic, leptonic, hadronic, rare and
forbidden decays of the charm quark, as well as \dmix\ mixing and CP violation
in the charm sector.  In each case we (i) review the motivation for
studying these processes, (ii) summarize the present status of theoretical
calculations in the SM and beyond, and (iii) examine the experimental
sensitivities available at a tau-charm factory.  Further details can be
found in Refs.\ [4,5].

\section{Semileptonic Decays of Charmed Hadrons}

Semileptonic decays occupy a fundamental place in the experimental study
of charm decays and in attempts to understand the underlying
dynamics associated with charmed hadrons.  They also play a critical role in
determining the elements of the CKM mixing matrix.
However, the extraction of the values for the CKM elements requires
knowledge of the form factors which parameterize the matrix elements of
the hadronic weak current.
Theoretical calculation of these form factors is a thorny problem,
particularly in the case of charm decays as they occur at an energy
scale which lies between the regimes where chiral perturbation theory and
heavy quark mass expansions (which are used to describe $K$ and $B$ decays,
respectively) are absolutely valid.  With the exception of lattice
calculations, the form factors cannot be calculated from first principles,
and thus model dependent estimates, usually combined with
approximate symmetry properties, must be employed.  It is therefore
crucial to measure the shape of the form factors in order to test the
theoretical models.  This would enhance our understanding of QCD, and enable
scaling of  the predictions to the
$B$ sector with better accuracy, as well as provide an improved
determination of the CKM matrix elements in the charm sector.

Before turning to a discussion of the form factors, we first note
that the CKM elements governing semileptonic charm decay, $V_{cd}$ and
$V_{cs}$, are surprisingly poorly determined at present.  The best
determination
of $V_{cd}$ is from charm production in neutrino scattering off valence
$d$-quarks, yielding the value\cite{pdg} $|V_{cd}|=0.204\pm 0.017$.
Extracting $V_{cs}$ from this method is plagued by uncertainties from
the estimates of the strange quark parton density and hence the semileptonic
decay $D^+\to \bar Ke^+\nu$ is used.  Comparison of the data with theory
and using
conservative assumptions in calculating the form factors yields\cite{pdg}
$|V_{cs}|=1.01\pm 0.18$.  The imposition of CKM unitarity for a three
generation SM constrains\cite{pdg} the values of these elements further to
$|V_{cd}|=0.221\pm 0.003$ and $|V_{cs}|=0.9745\pm 0.0007$.  We see that
these CKM elements are very well known if one {\it assumes} unitarity, but
that the direct experimental measurements are at the $10-20\%$ level.
There are many theories beyond the SM which preclude a three generation
unitary CKM matrix and they should be tested with a precision in the charm
sector that is equivalent to that of planned experiments
in the $B$ system.  For example, the comparison of the exclusive semileptonic
decays $D\to\pi\mu\nu$ and $D\to K\mu\nu$, where theoretical uncertainties
mostly cancel in taking the ratio,
would allow a determination of $|V_{cd}|/|V_{cs}|$
to $\sim 1\%$, which is comparable to the present level of precision on the
Cabibbo angle $\theta_c$.

Semileptonic rates are dominated by modes containing only one hadron in the
final state.  These are $K$ or $K^*$ for Cabibbo allowed decay, $\pi$
or $\rho$ for Cabibbo suppressed $D$ decay, and $\eta$, $\eta'$, $\phi$
for Cabibbo suppressed $D_s$ decay.  For the decay into a pseudoscalar
meson, $D\to P\ell\nu$, the hadronic matrix element can be written as
\begin{equation}
\langle P(k)|\bar q\gamma_\mu c|D(p)\rangle = f_+(q^2)(p+k)_\mu +
f_-(q^2)(p-k)_\mu \,,
\end{equation}
where $q^2=(p-k)^2$ is the momentum transfer and $\bar q=\bar s,\bar d$\
for $P=K,\pi$, respectively.  If the lepton mass is neglected, only $f_+$
contributes to the $q^2$ distribution with
\begin{equation}
{d\Gamma\over dq^2} = {G^2_F|V_{cs(d)}|^2k^3\over 24\pi^3}
|f_+(q^2)|^2 \,.
\end{equation}
We note that an accurate comparison of the rate for $D\to Ke\nu$ to that for
$D\to K\mu\nu$ would determine $f_-(q^2)$ and provide the best direct
measurement of $V_{cs}$.

Theoretical uncertainties are introduced by assuming a particular
$q^2$ dependence of the form factor.  It is usually parameterized using vector
meson dominance with
\begin{equation}
f_+(q^2)={ f_+(0)\over 1-q^2/m^{*2} } \,,
\end{equation}
where $m^*$ is the pole mass of the appropriate channel. This method
allows one to fit and determine the pole mass.  In heavy quark
effective theory (HQET), all decay modes are described in terms of a single
form
factor assuming an exponential $q^2$ dependence, $f_+(q^2)=f_+(0)\exp
(\alpha q^2)$.  This yields values
of $f_+(0)$ which are approximately $5\%$ smaller than those obtained
in the single pole ansatz.  However, HQET is less advantageous in $D$ decays
as the symmetry breaking corrections as expected to be large.  Various
theoretical predictions\cite{theoryff} (from quark model calculations,
QCD sum rules, and lattice QCD)
for the form factor in $D\to K\ell\nu$, as well as the average experimental
result\cite{pdg}, including
a recent determination by CLEO\cite{cleoff} from a measurement of
the $q^2$ distribution, is summarized in Table 2.
In the case of the monopole distribution, the pole mass is then measured
to be $m^*=(2.00\pm 0.12\pm 0.18)\gev$, which is consistent with the
$D^*_s$ mass.  For HQET, the fit value of $\alpha$ is $\alpha=(0.29\pm
0.04\pm 0.06)\gev^{-2}$, which should be compared with the predicted
value\cite{theoryff} of $\alpha=0.21\gev^{-2}$.

The various quark model predictions for the $q^2$ dependence are in rough
agreement with each
other as they all incorporate a recoil energy dependence that reflects
a fall-off with a typical hadronic scale of\ $\sim 1\gev$.
Large disagreements in the predictions will only occur either at low
energies, if there are poles close to the physical
region, or at recoil energies greater than 1 GeV.  Therefore charm meson
semileptonic decays do not fully probe the different aspects of the
models given the limited kinematic range.  Very precise measurements will
then be needed in order to discriminate amongst the models.

For semileptonic decay to a vector meson $V$, the hadronic matrix element
can be parameterized in terms of four form factors,

\begin{eqnarray}
\langle V(k)|\bar q\gamma_\mu c|D(p)\rangle & = & {2V(q^2)\over m_D+m_V}
\epsilon_{\mu\nu\alpha\beta}\epsilon^{*\nu}p^\alpha k^\beta \,, \\
\langle V(k)|\bar q\gamma_\mu\gamma_5 c|D(p)\rangle & = & i \left[
(m_D+m_V)A_1(q^2)\epsilon^*_\mu - {A_2(q^2)\over m_D+m_V}\epsilon^*
\cdot p(p+k)_\mu \right. \nonumber\\
& & \quad\quad  \left. +{2m_V\over q^2}A_3(q^2)\epsilon^*\cdot p(p-k)_\mu
\right] \,, \nonumber
\end{eqnarray}
where $\epsilon$ represents the polarization vector of $V$.
The contribution from $A_3(q^2)$ vanishes in the limit of massless leptons.
Here, branching fraction measurements alone do not allow the
determination of the form factors, and angular distributions must also be
measured in order to extract all three.
A summary of existing form factor measurements\cite{pdg} and various
theoretical predictions\cite{theoryff} is displayed in Table 2 for
comparison purposes.  We see that most calculations agree, within errors,
with the experimental
determination of $V(0)$, but fail to predict $A_1(0)$ and $A_2(0)$.
\begin{table}
\centering
\begin{tabular}{|c|c|c|c|c|} \hline\hline
Reference & $f_+(0)$ & $V(0)$ & $A_1(0)$ & $A_2(0)$ \\ \hline
Exp. Average & $0.75\pm 0.02\pm 0.02$ & $1.1\pm 0.2$ & $0.56\pm 0.04$
& $0.40\pm 0.08$ \\ \hline
Theory & & & & \\ \hline
ISGW & 0.82 & 1.1 & 0.8  & 0.8 \\
BSW  & 0.76 & 1.3 & 0.88 & 1.2 \\
AW   & 0.7  & 1.5 & 0.8  & 0.6 \\
BKS  & $0.90\pm 0.08\pm 0.21$ & $1.4\pm 0.5\pm 0.5$ & $0.8\pm 0.1\pm 0.3$
& $0.6\pm 0.1\pm 0.2$ \\
LMMS & $0.63\pm 0.08$ & $0.9\pm 0.1$ & $0.53\pm 0.03$ & $0.2\pm 0.2$ \\
BBD  & 0.60 & 1.1 & 0.5 & 0.6 \\ \hline\hline
\end{tabular}
\caption{Measured and predicted form factors in
$D\to K\ell\nu$ and $D\to K^*\ell\nu$.}
\end{table}

Theoretical predictions\cite{theoryff} for the ratio of vector to
pseudoscalar decay modes are compared to the experimental averaged
result\cite{pdg} in Table 3.  We see that the QCD sum rule approach
(BBD) obtains the best agreement with experiment, but also has large
errors.  The agreement with the $f_+(0)$ measurements seen in
Table 2 points to the $K^*$ mode as the problem, which is also in accord
with our conclusions on the axial vector form factors in Table 2.  Since we
seem to lack a clear theoretical understanding of these decays, accurate
experimental information in several modes is vital.
\begin{table}
\centering
\begin{tabular}{|c|c|} \hline\hline
Reference & $\Gamma(D\to K^*\ell\nu)/\Gamma(D\to K\ell\nu)$ \\ \hline
Experiment & $0.58\pm 0.06$  \\ \hline
ISGW & 1.14 \\
BSW  & 0.87 \\
AW   & 1.34 \\
BBD  & $0.6\pm 0.4$ \\ \hline\hline
\end{tabular}
\caption{Theoretical predictions and experimental average for the ratio
$\Gamma(D\to K^*\ell\nu)/\Gamma(D\to K\ell\nu$.}
\end{table}

There are two recent experimental measurements of the vector and axial-vector
form factors in $D_s$ decays.  They constitute a further opportunity to
examine the discrepancies between theory and experiment which appear to
exist in $D$ decay.  Table 4 summarizes the measured\cite{pdg} and
predicted\cite{theoryff}
form factors in the decay $D_s\to\phi\ell\nu$.  We see that
theory fails to provide the correct value for the ratio $A_2(0)/A_1(0)$,
just as in the $K^*$ mode.  It would be of interest to investigate the process
$D_s\to (\eta+\eta')\ell\nu$ to see if the
disagreement between theory and experiment persists only in the vector mode
as in the case of $D$ decays.
\begin{table}
\centering
\begin{tabular}{|c|c|c|} \hline\hline
Reference & $V(0)/A_1(0)$ & $A_2(0)/A_1(0)$ \\ \hline
Exp. Average & $2.0\pm 0.7$ & $ 1.8\pm 0.5$ \\ \hline
SI   & 1.85 & 1.21 \\
BKS  & $2.00\pm 0.19\pm 0.23$ & $0.78\pm 0.08\pm 0.15$ \\
LMMS & $1.65\pm 0.2$ & $0.33\pm 0.36$ \\ \hline\hline
\end{tabular}
\caption{Measured and calculated form factors in $D_s\to\phi\ell\nu$.}
\end{table}

Charmed baryon semileptonic decays provide yet another opportunity to test
theoretical calculations.  For example, a measurement of
$\Lambda_c^+\to\Lambda\ell\nu$,
which is expected to dominate the $\Lambda_c$ semileptonic decays, would
yield information on heavy quark predictions that relate these decays
to $D\to X_s\ell\nu$.  Measurement of the $\Lambda_c$ and $\Xi_c$ branching
factions would supply essential input to the analysis of the charm
content in $B$ decays.

A study of inclusive semileptonic decays would provide an additional test of
HQET, as well as allowing for the extraction of a value for the charm quark
mass.  The results of a recent HQET analysis\cite{bigi} from present data on
charm semileptonic decays yield $m_c=1.57\pm 0.03\pm
0.05\gev$, where the errors represent the experimental and theoretical
uncertainties, respectively.  Another important question to be addressed is
whether the sum of the exclusive rates saturate the inclusive rate.

With adequate statistics, the exclusive semileptonic rates as well as the $q^2$
dependences and relative normalization of the form factors can be measured at
a \tcf.  The ability to make efficient measurements over the full kinematic
range is essential.  The expected number\cite{rafe} of detected exclusive
modes per year at a \tcf\ lies in the range $10^4 - 10^5$ for the various
channels in $D^0, D^+$ decay and $10^3 - 10^4$ for $D_s$.  Figure 1 from
Ref.\ [10]
displays the use of tagging, kinematics, and calorimetry to isolate
Cabbibo allowed and suppressed decays by the method of reconstructing
the missing mass.  This analysis shows that
the events are expected to be essentially
background free.  A \tcf\ could provide an exhaustive and accurate data base,
with which the various theoretical models could be compared, and
allows for the reliable determination of the CKM elements governing
charm-quark decay.

\section{Leptonic Decays of Charmed Mesons}

The SM transition rate for the purely leptonic decay of a pseudoscalar charm
meson is
\begin{equation}
\Gamma(D^+_{(q)}\to\ell^+\nu_\ell)={G^2_F\over 8\pi}
f_{D_{(q)}}|V_{cq}|^2m_{D_{(q)}}m^2_\ell\left( 1-{m^2_\ell\over m^2_{D_{(q)}}}
\right)^2 \,,
\end{equation}
with $q=d, s$ and $f_{D_(q)}$ is the weak decay constant defined as usual by
\begin{equation}
\langle 0|\bar q\gamma_\mu\gamma_5 c| D_{(q)}({\bf p})\rangle = if_{D_{(q)}}
p_\mu \,,
\end{equation}
where $f_\pi=131\mev$ in this normalization.  The resulting branching
fractions are small due to the helicity suppression and are listed in
Table 5 using the central values of the CKM parameters given in Ref.\ [6]
and assuming $f_D=200\mev$ and $f_{D_s}=230\mev$.
\begin{table}
\centering
\begin{tabular}{|c|c|c|} \hline\hline
Meson & $\mu^+\nu_\mu$ & $\tau^+\nu_\tau$   \\ \hline
$D^+$   & $3.52\times 10^{-4}$ & $9.34\times 10^{-4}$ \\
$D^+_s$ & $4.21\times 10^{-3}$ & $4.11\times 10^{-2}$ \\ \hline\hline
\end{tabular}
\caption{SM branching fractions
for the leptonic decay modes, assuming $f_D=200\mev$ and $f_{D_s}=230\mev$.}
\end{table}

Assuming that the CKM matrix elements are well-known, the leptonic decays can
provide important information on the value of the
pseudoscalar decay constants.  Precise measurements
of these constants are essential for the study of \dmix\ mixing, CP
violation in the charm sector, and non-leptonic decays.  They would also
test the accuracy of QCD calculational techniques and provide a means
for more precise predictions when extrapolating to the $B$ sector.
The existing upper limit for $f_D$ is $f_D<
290\mev$, and is derived from the $90\%$ C.L. bound\cite{markthree}
$B(D^+\to\mu^+\nu_\mu)<7.2\times 10^{-4}$.  CLEO has
recently observed\cite{cleofds} the process
$D_s^{*+}\to D_s^+\gamma\to\mu\nu\gamma$ by examining the mass difference
$\delta M\equiv M_{\mu\nu\gamma}-M_{\mu\nu}$ and have obtained
\begin{equation}
{\Gamma(D_s^+\to\mu^+\nu)\over\Gamma(D_s^+\to\phi\pi^+)}=
0.235\pm 0.045\pm 0.063 \,.
\end{equation}
Using $\Gamma(D_s^+\to\phi\pi^+)=3.7\pm 1.2 \%$ they find $f_{D_s}=337\pm
34\pm 45\pm 54\mev$ where the last error reflects the uncertainty in the
$\phi\pi^+$ branching fraction.  An emulsion experiment has measured\cite{aoki}
$f_{D_s}=232\pm 45\pm 52 \mev$.
The BES Collaboration has also recently
reported\cite{besfds} the observation of three candidate events in
$e^+e^-\to D_s^+D_s^-$ with the subsequent decay $D_s\to\mu\nu$ yielding
$f_{D_s}=434^{+153~~+35}_{-133~~-33}\mev$.  The errors are expected to improve
once more statistics are obtained.

A variety of theoretical techniques have been employed to estimate the value
of $f_D$ and $f_{D_s}$.
Lattice QCD studies\cite{soni} calculate these quantities in the quenched
approximation through a procedure that interpolates between the Wilson
fermion scheme and the static approximation.
The non-relativistic quark model is used
to relate the decay constant to the meson wave function at the origin, $f_M=
\sqrt{12/M_M}\, |\psi(0)|$, which is then inferred from isospin mass splitting
of heavy mesons\cite{jim}.  Another approach uses QCD sum rules\cite{sumrule}.
For each of these calculational methods, the resulting ranges for the values
of the pseudoscalar decay constants are presented in Table 6.  Given the large
errors, the results of these three approaches are roughly consistent.
\begin{table}
\centering
\begin{tabular}{|c|c|c|c|} \hline\hline
Decay Constant & Lattice & Quark Model & Sum Rule \\ \hline
$f_D$     & $208(9)\pm 35\pm 12$ & $207\pm 60$ & 170 -- 235 \\
$f_{D_s}$ & $230(7)\pm 30\pm 18$ & $259\pm 74$ & 204 -- 270 \\
$f_{D_s}/f_D$ & $1.10(2)\pm 0.02\pm 0.02\pm 0.03$ & 1.25 & $1.21\pm 0.06$
\\ \hline\hline
\end{tabular}
\caption{Theoretical estimates of the weak decay constants in units of MeV,
(taking $m_c=1.3\gev$ in the sum rule approach).}
\end{table}
The theoretical uncertainties associated with the ratio $f_{D_s}/f_D$
are much smaller, as
this ratio should deviate from unity only in the presence of broken SU(3)
flavor symmetry.  The magnitude of such flavor violating effects can
be determined by the measurement of $B(D^+_s\to\mu^+\nu_\mu)/B(D^+\to
\mu^+\nu_\mu)$.  Obtaining precise values for these decay constants, both
theoretically and experimentally,
would provide valuable input for a wide range of
phenomenological applications.

Non-SM contributions may affect the purely leptonic decays.  Signatures for
new physics include the measurement of non-SM values for the absolute
branching ratios, or the observation of a deviation from the SM prediction
\begin{equation}
{B(D^+_{(s)}\to \mu^+\nu_\mu)\over B(D^+_{(s)}\to\tau^+\nu_\tau)}
={ {m^2_\mu\left( 1- m^2_\mu/ m^2_{D_{(s)}}\right)^2 }\over {m^2_\tau
\left( 1- m^2_\tau/ m^2_{D_{(s)}}\right)^2 }} \,.
\end{equation}
This ratio is sensitive to violations of $\mu-\tau$ universality.

As an example, we consider the case where the SM Higgs sector is
enlarged by an additional Higgs doublet.  These models generate
important contributions\cite{btaunu} to the decay $B\to\tau\nu_\tau$ and
it is instructive to examine their effects in the charm sector.  Two such
models, which naturally avoid tree-level flavor changing neutral currents,
are Model I, where one doublet ($\phi_2$) generates masses for all
fermions and the second ($\phi_1$) decouples from the fermion sector, and
Model II, where $\phi_2$ gives mass to the up-type quarks, while the down-type
quarks and charged leptons receive their mass from $\phi_1$.  Each doublet
receives a vacuum expectation value $v_i$, subject to the constraint that
$v_1^2+v_2^2=v^2_{\rm SM}$.  The charged Higgs boson present in these models
will mediate the leptonic decay through an effective four-Fermi interaction,
similar to that of the SM $W$-boson.  The $H^\pm$ interactions with
the fermion sector are governed by the Lagrangian
\begin{eqnarray}
{\cal L} & = & {g\over 2\sqrt 2 M_W}H^\pm[V_{ij}m_{u_i}A_u\bar
u_i(1-\gamma_5)d_j+V_{ij}m_{d_j}A_d\bar u_i(1+\gamma_5)d_j \nonumber \\
& & \quad\quad\quad\quad m_\ell A_\ell\bar\nu_\ell(1+\gamma_5)\ell]+h.c. \,,
\end{eqnarray}
with $A_u=\cot\beta$ in both models and $A_d=A_\ell=-\cot\beta(\tan\beta)$ in
Model I(II), where $\tan\beta\equiv v_2/v_1$.  In Models I and II, we obtain
the result
\begin{equation}
B(D^+\to\ell^+\nu_\ell)  =  B_{\rm SM}\left( 1+ {m_D^2\over m^2_{H^\pm}}
\right)^2 \,,
\end{equation}
where in Model II the $D_s^+$ decay receives an additional modification
\begin{equation}
B(D^+_s\to\ell^+\nu_\ell)  =  B_{\rm SM}\left[ 1+ {m_{D_s}^2\over m^2_{H^\pm}}
\left(1- \tan^2\beta{m_s\over m_c}\right)\right]^2 \,.
\end{equation}
In this case, we see that the effect of the $H^\pm$ exchange is independent
of the leptonic final state and the above prediction for the ratio in Eq. (8)
is unchanged.  This is because the $H^\pm$ contribution is proportional
to the charged lepton mass, which is then a common factor with the SM helicity
suppressed term.
However, the absolute branching fractions can be modified; this effect
is negligible in the decay $D^+\to\ell^+\nu_\ell$, but could be of
order a few percent in $D^+_s$ decay if $\tan\beta$ is very large.

The detection of $D^+, D^+_s\to\mu^+\nu$ decays is straightforward
with single tagged event samples at a
\tcf\ allowing the measurement of the weak decay constants at the
$1\%$ level\cite{kim}.  The $\tau^+\nu$ modes can also be observed via
the subsequent decays $\tau\to\pi\nu,\ell\nu\nu$ by relying on
tagging, hermiticity, and background measurements.  The tagged event
samples are used to provide a constrained fit for the neutrino mass
as well as to suppress backgrounds.  Figure 2 from Ref.\ [19]
shows that the missing mass spectrum cleanly separates the
signal from background in all cases.

\section{Nonleptonic Decays of Charm Hadrons}

Nonleptonic charm decays provide another opportunity to test our
theoretical understanding of heavy flavor decays.  They occur in a
regime where non-perturbative effects are more (less) important than
in $B$ ($K$) decays.  The thorough investigation of the rich variety
of decay processes available would unravel these QCD effects and
would build a foundation for a more reliable
extrapolation to the underlying dynamics in the $B$ system.  Several types
of processes mediate nonleptonic charm decays, \eg, external spectator,
color suppressed or internal spectator, weak annihilation, exchange,
penguin, and mixing.  Disentangling the underlying quark processes and
determining their relative strengths is difficult and will require an increase
of $2-3$ orders of magnitude over current statistics, but will generate a
detailed understanding of strong interaction effects.

The absolute branching fractions of hadronic charm decays need to
be measured with better precision.  The present sensitivity of
measurements of $D^0$ and $D^+$ decay is at the $10-20\%$ level, while
the situation for $D_s$ is even more uncertain as the absolute scale of its
branching fractions is unknown.  Most $D_s$ decays are measured relative
to the $\phi\pi^+$ mode, which is determined from the ratio $\Gamma(D_s^+
\to\phi e^+\nu)/\Gamma(D_s^+\to\phi\pi^+)$ combined with estimates of
the $D_s^+$ cross section, theoretical predictions relating $D^+$ and
$D_s^+$ decays, and fragmentation assumptions.  A direct determination
of the $\phi\pi^+$ branching fraction could thus shift the values of
the remaining branching fractions.  Very few of the rates for charm baryon
decay (particularly for $\Xi_c$ and $\Omega_c$) have been measured.
Accurate data on these modes would provide invaluable assistance in the
determination of the charm content in $B$ decays, CP violating asymmetries,
and in fragmentation studies.

The \tcf\ is well suited\cite{rafe} for these measurements.  The increase in
statistics and the efficient single and double tagging techniques
allows the determination of the absolute branching fractions for
$D^0$ and $D^+$ at the $1\%$ level, limited only by systematics.
Approximately $5\times 10^5$ single $D_s$ tags can be reconstructed in
one year at $10^{33}\, {\rm cm}^{-2}\, {\rm sec}^{-1}$ using the $\phi\pi^+\,,
S^*\pi^+\,, \bar K^0K^+$ and $\bar K^{*0}K^+$ decay modes, while the
number of double tags from pairing these channels is $4950\times B(\phi
\pi^+)$.  This yields a\ $\sim 3\%$ sensitivity to the absolute scale of
$D_s$ decays.  Similar absolute measurements of charm baryon decays could
also be performed.

$D$ mesons are sufficiently heavy for many body final states to play a
prominent role in their nonleptonic modes, however, $D^0$ and $D^+$
decays presently appear to be dominated by two body modes.  In this
case, most of the theoretical work is based on factorization\cite{bsw}
or QCD sum rule\cite{bs} approaches.  The latter technique yields
approximately model independent predictions for the amplitudes of four
decay types $D\,, D_s\to PP, PV$,  where $P(V)$ is a pseudoscalar (vector)
meson built of light quarks, but the numerical results depend on
the assumption of $SU(3)$ flavor symmetry.  The
former method is based on $PP, VP, VV$, and $AP$ final states (where $A$
represents an axial-vector meson) and assumes, (i) an amplitude which is
expressed as the product of two current matrix elements, \ie,
\begin{equation}
\langle M_1M_2|J^\mu J_\mu|D\rangle \simeq \langle M_1|J^\mu|0\rangle
\cdot\langle M_2|J_\mu|D\rangle \,,
\end{equation}
(ii) two fit parameters, $a_1$ and $a_2$, which occur in the
nonleptonic weak Hamiltonian
\begin{equation}
{\cal H}^{eff}=-{G_F\over\sqrt 2}:[a_1(\bar ud')(\bar s'c)+
a_2(\bar s'd')(\bar uc)]: \,,
\end{equation}
where the colons denote Wick ordering and $d', s'$ are Cabbibo rotated,
(iii) model dependencies for the hadronic wavefunctions, and (iv) that
the contributions from penguin amplitudes and final state interactions are
negligible.
This model currently provides a reasonable description for $D^0$ and
$D^+$ decay modes, but may not hold up once more accurate data is obtained.
The systematic study and comparison of $D$ and $B$ decays will help clarify
some of the unresolved issues and inherent model dependencies associated with
factorization.

The situation with $D_s$ decays is not so well described.  The anticipated
strength of the weak annihilation process has yet to be confirmed and it is
possible that the pattern of $D_s$ decays may not mimic those of $D^0$
and $D^+$.  It has been suggested\cite{bs} that the $D_s$ may have an
enhanced non-resonant width into multi-particle final states.
In this case, the constraints of production at threshold help to
reconstruct these complicated many body modes.  An reconstruction example
from Ref.\ [3] at $\sqrt s=
4.028\gev$ for $D_s\to\eta\pi^+\pi^+\pi^-$ with the subsequent decays
$\eta\to\pi^+\pi^-\pi^0$ and $\pi^0\to\gamma\gamma$ is depicted in Fig. 3.

Doubly suppressed Cabibbo decays, having branching fractions of order
$\tan^4\theta_c$ relative to Cabbibo allowed decays, also test our
understanding of the hadronic weak current.  These modes are best measured
by tagging in $D^+$ decays, since there is no confusion with the possible
mixing component which may be present in $D^0$ modes.  A handful of
events in $D^+\to K^+K^+K^-, \phi K^+$
and $D^0\to K^+\pi^-$ have been observed\cite{pdg}.
It is expected\cite{rafe}
that a few hundred such events can be reconstructed at the \tcf\ and that a
signal sensitivity at the $\sim 10\%$ level can be achieved.

\section{Rare and Forbidden Decays of Charm Mesons }

Flavor changing neutral current (FCNC) decays only occur at the loop
level in the SM.  Due to the effectiveness of the GIM mechanism and the
small masses of the quarks which participate inside the loops,
short distance SM contributions to rare charm decays are very small.
Most reactions are thus dominated by long distance effects which are
difficult to reliably calculate.  However, a recent
investigation\cite{charmcrew} of such effects indicates
that there is a window for the clean observation of new physics in some
interactions.  In fact, it is precisely because the SM
FCNC rates are so small, that charm provides an important opportunity
to discover new effects, and offers
a detailed test of the SM in the up-quark sector.

FCNC decays of the $D$ meson include the processes $D^0\to\ell^+\ell^-,
\gamma\gamma$, and $D\to X+\ell^+\ell^-, X+\nu\bar\nu, X+\gamma$, with
$\ell=e, \mu$.  They proceed via electromagnetic or weak penguin
diagrams as well as receiving contributions from box diagrams in some
cases.  The calculation of the SM short distance rates for these processes is
straightforward and the transition amplitudes and standard loop integrals,
which are categorized in Ref.\ [22] for rare $K$ decays, are easily
converted to the $D$ system.  The loop integrals relevant for
$D^0\to\gamma\gamma$ may be found in Ref.\ [23].  Employing the GIM
mechanism results in a general expression for the loop integrals which
can be written as
\begin{equation}
A=V_{cs}V^*_{us}[F(x_s)-F(x_d)]+V_{cb}V^*_{ub}[F(x_b)-F(x_d)] \,,
\end{equation}
with $x_i\equiv m^2_i/M_W^2$ and $F(x_d)$ usually being neglected (except in
the $2\gamma$ case).  The $s$- and $b$-quark contributions are roughly
equal as the larger CKM factors compensate for the small strange quark mass.
The values of the resulting inclusive short distance branching fractions,
before
QCD corrections are applied, are
shown in Table 7, along with the current experimental bounds\cite{pdg,raredk}.
The corresponding
exclusive rates are typically an order of magnitude less than the inclusive
case.  We note that the
transition $D^0\to\ell^+\ell^-,$ is helicity suppressed and hence has the
smallest branching fraction.  The range given for this branching fraction,
$(1-20)\times 10^{-19}$, indicates the effect of varying the parameters
in the ranges $f_D=0.15-0.25\gev$ and $m_s=0.15-0.40\gev$.
\begin{table}
\centering
\begin{tabular}{|l|c|c|c|} \hline\hline
Decay Mode & Experimental Limit & $B_{S.D.}$ & $B_{L.D.}$ \\ \hline
$D^0\to\mu^+\mu^-$ & $<1.1\times 10^{-5}$ & $(1-20)\times 10^{-19}$ &
$<3\times 10^{-15}$ \\
$D^0\to e^+e^-$ & $<1.3\times 10^{-4}$ &  & \\
$D^0\to\mu^\pm e^\mp$ & $<1.0\times 10^{-4}$ & $0$ & $0$ \\ \hline
$D^0\to\gamma\gamma$ & --- & $10^{-16}$ & $<3\times 10^{-9}$ \\ \hline
$D\to X_u+\gamma$ & & $1.4\times 10^{-17}$ & \\
$D^0\to\rho^0\gamma$ & $<1.4\times 10^{-4}$ & & $<2\times 10^{-5}$ \\
$D^0\to\phi^0\gamma$ & $<2.0\times 10^{-4}$ & & $<10^{-4}$ \\
$D^+\to\rho^+\gamma$ & --- & & $<2\times 10^{-4}$ \\
$D^+\to\bar K^{*+}\gamma$ & --- & & $3\times 10^{-7}$ \\
$D^0\to\bar K^{*0}\gamma$ & ---& & $1.6\times 10^{-4}$ \\ \hline
$D\to X_u+\ell^+\ell^-$ & & $4\times 10^{-9}$ & \\
$D^0\to\pi^0\mu\mu$ & $<1.7\times 10^{-4}$ & & \\
$D^0\to\bar K^0 ee/\mu\mu$ & $<17.0/2.5\times 10^{-4}$ & &
$<2\times 10^{-15}$ \\
$D^0\to\rho^0 ee/\mu\mu$ & $<2.4/4.5\times 10^{-4}$ & & \\
$D^+\to\pi^+ee/\mu\mu$ & $<250/4.6\times 10^{-5}$ & few$\times 10^{-10}$ &
$<10^{-8}$ \\
$D^+\to K^+ee/\mu\mu$ & $<480/8.5\times 10^{-5}$ & & $<10^{-15}$ \\
$D^+\to\rho^+\mu\mu$ & $<5.8\times 10^{-4}$ & & \\
\hline
$D^0\to X_u+\nu\bar\nu$ & & $2.0\times 10^{-15}$ & \\
$D^0\to\pi^0\nu\bar\nu$ & --- & $4.9\times 10^{-16}$ & $<6\times 10^{-16}$ \\
$D^0\to\bar K^0\nu\bar\nu$ & --- & & $<10^{-12}$ \\
$D^+\to X_u+\nu\bar\nu$ & --- & $4.5\times 10^{-15}$ & \\
$D^+\to\pi^+\nu\bar\nu$ & --- & $3.9\times 10^{-16}$ & $<8\times 10^{-16}$ \\
$D^+\to K^+\nu\bar\nu$ & --- & & $<10^{-14}$ \\ \hline\hline
\end{tabular}
\caption{Standard Model predictions for the branching fractions due to short
and long distance contributions for various rare $D$ meson decays. Also
shown are the current experimental limits.}
\end{table}

The calculation of the long distance branching fractions are plagued with
the usual hadronic uncertainties and the estimates listed in the table convey
an upper limit on the size of these effects rather than an actual value.
These estimates have been computed by considering various intermediate
particle states (\eg, $\pi, K, \bar K, \eta, \eta', \pi\pi,\ {\rm or}\ K\bar
K$)
and inserting the known rates for the decay of the intermediate particles
into the final state of interest.  In all cases we see that the long
distance contributions overwhelm those from SM short distance physics

The radiative decays, $D\to X+\gamma$, merit further discussion.  One of the
goals of a high luminosity $B$ physics program is to extract the ratio of CKM
elements $|V_{td}|/|V_{ts}|$ from a measurement of $B(B\to\rho/\omega+\gamma)/
B(B\to K^*\gamma)$.  CLEO has placed\cite{cleobsg} a bound on this ratio of
branching fractions of $<0.34$, while observing the decay $B\to K^*\gamma$
with a branching fraction of $(4.5\pm 1.5\pm 0.9)\times 10^{-5}$.  This yields
a very loose constraint on the above ratio of CKM elements.  This
method of determining the ratio of CKM elements depends critically
on the assumption that these exclusive decay modes are dominated by short
distance penguin transitions.  If this assumption is false, and the long
distance contributions to these decays were found to be large, this
technique would be invalidated.  Unfortunately, the theoretical uncertainties
associated with computing the long distance contributions are
sizeable\cite{sandip}.
Where separation of the two types of contributions is somewhat difficult in
the $B$ sector, radiative charm decays provide an excellent testing ground.
In this
case it should be possible to separate out the inclusive penguin transitions
$c\to u\gamma$, and determine the rate of the long distance reactions which
are expected to dominate.  For example, the penguin transitions do not
contribute to $D^0\to\bar K^{*0}\gamma$ and it would be a direct measurement
of the non-perturbative effects.  Before
QCD corrections are applied, the short distance inclusive rate is very
small, $B(c\to u\gamma)=1.4\times 10^{-17}$; however, the QCD corrections
{\it greatly} enhance this rate\cite{radcharm}.  These corrections are
calculated via an operator product expansion, where the effective Hamiltonian
is evolved at leading logarithmic order from the electroweak scale to down to
the charm quark scale by
the Renormalization Group Equations.  This procedure mirrors that used for
$b\to s\gamma$, and results in $B(c\to u\gamma)=(1.1 - 2.3)\times 10^{-5}$,
where the lower(upper) value corresponds to taking the scale of the decay to
be $2m_c(m_c)$.  (We note that these radiative branching fractions have been
scaled to semileptonic charm decay in order to reduce the CKM and $m_c$
uncertainties.)  In this case, the rate is given almost entirely by
operator mixing!  The penguin contributions to the exclusive channels would
then be typically of order $10^{-6}$, which is still smaller than the long
distance estimates in Table 7.  Observation of
\begin{equation}
{B(D^0\to\rho^0\gamma)\over B(D^0\to\bar K^{*0}\gamma)}\neq \tan^2\theta_c
\end{equation}
at the few percent level would be a test of the perturbative QCD corrections,
or it could be a signal for new physics.
We note that the predicted values of the branching fractions for radiative
charm decays are well within reach of the \tcf.

Lepton flavor violating decays, \eg, $D^0\to\mu^\pm e^\mp$ and
$D\to X+\mu^\pm e^\mp$, are strictly forbidden in the SM with massless
neutrinos.  In a model with massive non-degenerate neutrinos and
non-vanishing neutrino mixings, such as in four generation models,
$D^0\to\mu^\pm e^\mp$ would be mediated by box diagrams with the massive
neutrinos being exchanged internally.  LEP data restricts\cite{neutr} heavy
neutrino mixing with $e$ and $\mu$ to be $|U_{Ne}U^*_{N\mu}|^2<7\times
10^{-6}$ for a neutrino with mass $m_N>45\gev$.  Consistency with
this bound constrains the branching fraction to be $B(D^0\to\mu^\pm
e^\mp)<6\times 10^{-22}$.  This same results
also holds for a heavy singlet neutrino
which is not accompanied by a charged lepton.  The observation
of this decay at a larger rate than the above bound
would be a clear signal for the existence of
a different class of models with new physics.

Examining Table 7, we see that there is a large window of opportunity
to discover the existence of new physics in rare charm decays.  Although
the SM short distance contributions are completely dominated by the
long distance effects, there are some modes where the size of the two
contributions are not that far apart.  The observation of any of
these modes at a larger rate than what is predicted from long
distance interactions would provide a clear signal for new physics.
To demonstrate the magnitude of enhancements that are possible in
theories beyond the SM, we consider two examples (i) a heavy $Q=-1/3$ quark
contributing to $D\to X+\ell^+\ell^-$ and (ii) leptoquark exchange
mediating $D^0\to \mu^\pm e^\mp$.  In the first case, a heavy $Q=-1/3$
quark may be present, \eg, as an iso-doublet fourth generation $b'$ quark,
or as a singlet quark in $E_6$ grand unified theories.
The current bound\cite{pdg} on the mass of such an object
is $m_{b'}>85\gev$, assuming that it decays via charged current interactions.
The heavy quark will then participate inside the penguin and box diagrams
which mediate the decay and result in the branching fractions presented
in Fig. 4.  The branching fractions are displayed as functions of
the overall CKM mixing factor for several values of the heavy quark mass,
and we see that a sizeable
enhancement is possible for large values of the mixing.
A naive estimate in the four generation SM yields\cite{pdg} the restrictions
$|V_{cb'}|<0.571$ and $|V_{ub'}|<0.078$.  In the second example of new
physics contributions we consider leptoquark bosons.  Leptoquarks are
color triplet particles which couple to a lepton-quark pair and are
naturally present in many theories beyond the SM which relate leptons and
quarks at a more fundamental level.  We parameterize their {\it a priori}
couplings as $\lambda^2_{\ell q}/4\pi=F_{\ell q}\alpha$.  Leptoquarks can
mediate $D^0\to \mu^\pm e^\mp$ by tree-level exchange, however their
contributions are suppressed by angular momentum conservation.  From the
limit $B(D^0\to \mu^\pm e^\mp)<10^{-4}$,
Davidson \etal\cite{sacha} find
\begin{equation}
\sqrt{F_{eu}F_{\mu c}} < 4\times 10^{-3} {\alpha\over 4\pi}
\left[{m_{LQ}\over 100\gev}\right]^2 \,,
\end{equation}
where $m_{LQ}$ represents the leptoquark mass.

Detector studies for the observation of loop-induced decays with leptonic
final states $\ell^+\ell^-$ and $X+\ell^+\ell^-$ were performed in
Ref.\ [30], where it was
estimated that the level of background was $\leq 1-10$ events and that
a \tcf\ could search for these decays with branching fractions down to
$(3-20)\times 10^{-8}$.  This is an improvement of $3-5$ orders of magnitude
over present limits and puts a \tcf\ within striking range for some of
these modes.  The radiative $D$ decays will be able to be studied in
detail at a \tcf.

\section{$D^0-\bar D^0$ Mixing}

Currently, the best limits\cite{pdg} on \dmix\ mixing are from fixed target
experiments, with $x_D\equiv\Delta m_D/\Gamma<0.083$ (where $\Delta
m_D=m_2-m_1$
is the mass difference), yielding $\dm<1.3\times 10^{-13}\gev$.
The bound on the ratio of wrong-sign to right-sign final
states is $r_D\equiv\Gamma(D^0\to\ell^-X)/\Gamma(D^0\to\ell^+X)<3.7\times
10^{-3}$, where
\begin{equation}
r_D\approx {1\over 2}\left[ \left( {\Delta m_D\over\Gamma}\right)^2 +
\left( {\Delta\Gamma\over 2\Gamma}\right)^2\right] \,,
\end{equation}
in the limit $\Delta m_D/\Gamma, \Delta\Gamma/\Gamma\ll 1$.

The short distance SM contributions to \dm\ proceed through a $W$ box diagram
with internal $d,s,b$-quarks.  In this case the external momentum, which is
of order $m_c$, is communicated to the light quarks in the loop and
can not be neglected.  The effective Hamiltonian is
\begin{equation}
{\cal H}^{\Delta c=2}_{eff} = {G_F\alpha\over 8\sqrt 2\pi x_w}\left[
|V_{cs}V^*_{us}|^2 \left(I_1^s {\cal O}-m_c^2I_2^s {\cal O'}\right)+
|V_{cb}V^*_{ub}|^2\left( I_3^b {\cal O}-m_c^2I_4^b {\cal O'}\right) \right] \,,
\end{equation}
where the $I_j^{q}$ represent integrals\cite{datta} that are functions of
$m_{q}^2/M_W^2$ and $m_{q}^2/m_c^2$, and ${\cal O}=[\bar u\gamma_\mu
(1-\gamma_5)c]^2$ is the usual mixing operator while ${\cal O'}=[
\bar u(1+\gamma_5)c]^2$ arises in the case of non-vanishing external
momentum.  The numerical value of the short distance contribution is
$\dm\sim 5\times 10^{-18}$ GeV (taking $f_D=200\mev$).  The long distance
contributions have been computed via two different techniques: (i) the
intermediate particle dispersive approach
(using current data on the intermediate states) yields\cite{charmcrew,gusto}
$\dm\sim 10^{-4}\Gamma\simeq 10^{-16}\gev$, and (ii) heavy quark
effective theory which results\cite{hqet} in $\dm\sim (1-2)\times 10^{-5}
\Gamma\simeq 10^{-17}\gev$.  Clearly, the SM predictions lie far below the
present experimental sensitivity!

One reason the SM expectations for \dmix\ mixing are so small is that there
are no heavy particles participating in the box diagram to enhance the rate.
Hence the first extension to the SM that we consider is the
addition\cite{four} of a heavy $Q=-1/3$ quark.
We can now neglect the external momentum and \dm\ is given
by the usual expression\cite{inlim},
\begin{equation}
\dm={G_F^2M_W^2m_D\over 6\pi^2}f_D^2B_D|V_{cb'}V_{ub'}^*|^2F(m^2_{b'}/M_W^2)
\,.
\end{equation}
The value of \dm\ is displayed in this model in Fig.\ 5a as a function of the
overall CKM mixing factor for various values of the heavy quark mass.  We see
that \dm\ approaches the experimental bound for large values of the
mixing factor.

Another simple extension of the SM is to enlarge the Higgs sector by an
additional doublet as discussed in the above leptonic decay section.
First, we examine two-Higgs-doublet models which
avoid tree-level FCNC by introducing a global symmetry.
The expression for \dm\ in these models can be found in Ref.\ [35].
{}From the Lagrangian in Eq. (9) it is clear that Model I will only modify the
SM
result for \dm\ for very small values of $\tan\beta$, and this region is
already
excluded\cite{cleobsg,bhp} from existing data on
$b\to s\gamma$ and $B_d^0-\overline B_d^0$
mixing.  However, enhancements can occur in Model II for large values of
$\tan\beta$, as demonstrated in Fig.\ 5b.

Next we consider the case of extended Higgs sectors without natural flavor
conservation.  In these models the above requirement of a global symmetry
which restricts each fermion type to receive mass from only one doublet is
replaced\cite{fcnch} by approximate flavor symmetries which act on the
fermion sector.  The Yukawa couplings can then possess a structure which
reflects the observed fermion mass and mixing
hierarchy.  This allows the low-energy FCNC limits to be evaded as the
flavor changing couplings to the light fermions are small.  We employ the
Cheng-Sher ansatz\cite{fcnch}, where the flavor changing couplings of the
neutral Higgs are $\lambda_{h^0f_if_j}\approx (\sqrt 2G_F)^{1/2}
\sqrt{m_im_j}\Delta_{ij}$, with the $m_{i(j)}$ being the relevant fermion
masses and $\Delta_{ij}$ representing a combination of mixing angles.
$h^0$ can now contribute to \dm\ through tree-level exchange
as well as mediating \dmix\ mixing by $h^0$ and t-quark virtual
exchange in a box diagram.  These latter contributions only compete with those
from the tree-level process for large values of $\Delta_{ij}$.  In Fig. 6a we
show the constraints placed on the parameters of this model from the present
experimental bound on \dm\ for both the tree-level and box diagram
contributions.

The last contribution to \dmix\ mixing that we will discuss here is that
of scalar leptoquark bosons.
They participate in \dm\ via virtual exchange inside a box
diagram\cite{sacha}, together with a charged lepton or neutrino.
Assuming that there is no leptoquark-GIM mechanism, and taking both exchanged
leptons to be the same type, we obtain the restriction
\begin{equation}
{F_{\ell c}F_{\ell u}\over m^2_{LQ}} <
{196\pi^2 \dm\over (4\pi\alpha f_D)^2m_D} \,,
\end{equation}
where $F_{\ell q}$ is defined in the previous section.
The resulting bounds in the leptoquark coupling-mass plane are presented
in Fig. 6b.

Signatures for \dmix\ mixing include like-sign dileptons from $D^0\bar D^0
\to\ell^\pm\ell^\pm X$ or dual hadronic decays such as $D^0\bar D^0\to
K^\pm\pi^\mp K^\pm\pi^\mp$.  The hadronic signal can be cleanly separated
from double Cabbibo suppressed decays at a \tcf\ since quantum statistics
yield different correlations.  This is due to the fact that the double Cabbibo
suppressed modes can only contribute to the same final states when the
$D^0$ mesons are in a relative S-wave.  The interference with the doubly
suppressed modes can also be used\cite{bigitcf} to separate mixing which
originates in the mass difference \dm\ from that which arises in
the decay $\Delta\Gamma_D$.  An extensive study of the double
Cabbibo decays will result as a product of the search for \dmix\ mixing.
A thorough Monte Carlo simulation of \dmix\ mixing at a \tcf\ detector
has been performed in Ref.\ [38], where the available sensitivity for mixing
was found to be at the $10^{-5}$ level.

\section{CP Violation}

CP violation in the $Q=2/3$ quark sector is complimentary to that of the
$K$ and $B$ systems, but has yet to be explored.  The \tcf\ factory
could thus provide an important first opportunity to explore CP
violation in this sector.  In the SM, the CKM phase is responsible for
generating CP violation, and the resulting rates are small.  However, new
sources of CP violating phases could greatly enhance the rates thus
rendering CP violation in the charm system a sensitive probe for physics
beyond the SM.

CP violation requires the interference of at least two amplitudes with
non-vanishing relative phases.  This can occur indirectly via \dmix\
mixing, or directly via asymmetries induced in the decay amplitude, or
kinematically in final state distributions.  The first case corresponds to
the interference of a $D^0$ decaying to a final state $f$ at time $t$,
with a $D^0$, which mixes into a $\bar D^0$ and then decays to $f$
at time $t$.  This process is theoretically clean as the hadronic
uncertainties cancel in the asymmetry.  However, since \dm\ is extremely
small in the SM the induced CP violation is negligible.
If new physics were to enhance \dmix\ mixing, as seen to occur in the previous
section for some models, then this mechanism could yield sizeable
CP violating effects.

Direct CP violation in charm meson decays yields slightly more promising
results.  Here, only decay amplitudes with two separate weak phases and two
different strong phases will contribute.  This can be easily seen as follows.
Let us assume that the decay amplitude to final state $f$ has the form
\begin{equation}
A_f=A_1 e^{i\delta_1}+A_2 e^{i\delta_2} \,,
\end{equation}
with $A_{1,2}$ being the two amplitudes after the strong phases $\delta_{1,2}$
have been factored out.  For the CP conjugate amplitude, the weak phases
are conjugated, but the strong phases are not.  The CP asymmetry is then
given by
\begin{equation}
{|A_f|^2-|\bar A_{\bar f}|^2 \over |A_f|^2+|\bar A_{\bar f}|^2} =
{2\Im (A_1^*A_2)\sin(\delta_1-\delta_2)\over |A_1|^2+|A_2|^2+2\Re (A_1^*A_2)
\cos(\delta_1-\delta_2)} \,,
\end{equation}
which clearly vanishes if $A_{1,2}$ contain the same weak phase and if
$\delta_1=\delta_2$.  Before estimating the typical size of this asymmetry
in the SM, we first note that
in contrast to $B$ decays, the branching fractions for
the relevant modes, \ie, $\pi^+\pi^-, K^+K^-$, etc., are rather sizeable
in the charm system, and for once, the large effects of final state
interactions are welcomed!  Of course, this makes numerical
predictions difficult due to the hadronic uncertainties associated with
final state interaction phases.  All tree-level interactions in $D$ decays
contain the same weak phase, hence they must interfere with penguin
mediated processes in order to obtain the requisite two independent weak
phases.  Since Cabibbo favored modes are not mediated via penguin type
interactions, only Cabibbo suppressed decays exhibit direct CP violation.
Although the relative size of the tree and penguin contributions has not yet
been determined, an upper bound for the size of the CP asymmetry is
estimated\cite{gusto,bigitcf,bucella} to be $a_{CP}<10^{-3}$.  However,
there could be a large enhancement from the strong interactions  (which
could be provided by, \eg, nearby resonances) and typical results in
this case are $a_{CP}\sim {\rm few}\times 10^{-3}$.

It is possible to obtain two separate weak phases in tree-level
amplitudes in $D_s$ decays, \ie, via interference between the
spectator and annihilation diagrams.  If the annihilation processes
are not suppressed relative to the spectator case, then CP asymmetries
of order $10^{-3}$ are feasible.

Kinematic CP violation signals could occur, for example, in the decays
$D\to VV$, which are described by more than one amplitude.  Here,
it is possible to construct CP-odd triple product correlations between the
two polarizations and one of the momenta, \eg, $\langle k\cdot\epsilon_1
\times\epsilon_2\rangle$.  Since final state interactions can induce a
non-zero value of the triple product correlation and hence mask CP violation,
one must evaluate
\begin{equation}
N_f={N(k\cdot\epsilon_1\times\epsilon_2>0)-N(k\cdot\epsilon_1
\times\epsilon_2<0)\over N_{total} } \,,
\end{equation}
as well as the corresponding quantity for $N_{\bar f}$, which vanish if CP is
conserved.  If $N_f$ and $N_{\bar f}$ are determined to be non-zero then
there is a clear signal of CP violation.
Similar correlations have also been discussed\cite{cpsemi} for semileptonic
decays.

An interesting example of the potential size of CP violating effects
from new physics is that of left-right symmetric models\cite{orsay,dipper}.
In this case reasonably large values for CP asymmetries can be obtained
for the Cabbibo allowed decay modes.  This occurs due to the existence of
an additional amplitude from the $W_R$ exchange, which carries a
different weak phase from that of the $W_L$ mediated decay.  The estimated
values of the CP asymmetries in these models is of order 0.01.  Compared to
the vanishing
asymmetry obtained in the SM, this would provide a sizeable and clear
signature for new physics.

The experimental feasibility of discovering CP violation at a \tcf\ has been
examined in detail\cite{fry,bigitcf}, as in the case of \dmix\ mixing,
where Monte Carlo results indicate that the backgrounds to the two-body
hadronic decays are manageable.
These authors find that a \tcf\ should be able to probe direct CP
violating asymmetries in $D$ decay at the level of $10^{-3}$, which just
touches the range predicted within the SM.

\section{Summary}

In summary we see that
the  charm physics program at a Tau-Charm factory is diverse and robust.
We have seen that (i) it is possible to greatly improve the precision over
present data samples on the charm system.  Most decay modes as well as the
pseudoscalar decay constants and the CKM elements governing charm-quark decay
can be measured at a level of order $1\%$.  This would greatly enhance our
understanding of the underlying QCD interactions which govern these processes.
In some decay modes and other interactions the \tcf\
could provide the first opportunity for observation.
(ii) Loop-induced reactions could be probed at a
substantially increased level of sensitivity  allowing for a wide window
of opportunity to discover
new physics. (iii) CP violation could finally be explored in
the up-quark sector, where the capabilities of a \tcf\ in one year of running
at design luminosity just touches the range of SM predictions.  Several
years of running (or increased luminosity) could study CP violation at the SM
level.
This wide physics potential and the high precision
which can be achieved
would place measurements in the charm sector at an equal level with
those in the down-quark systems.

The \tcf\ has the potential to address fundamental questions of the SM
in a complimentary fashion to the on-going extensive studies of the
down-quark sector.

\vspace{1.0cm}
%
\def\MPL #1 #2 #3 {Mod.~Phys.~Lett.~{\bf#1},\ #2 (#3)}
\def\NPB #1 #2 #3 {Nucl.~Phys.~{\bf#1},\ #2 (#3)}
\def\PLB #1 #2 #3 {Phys.~Lett.~{\bf#1},\ #2 (#3)}
\def\PR #1 #2 #3 {Phys.~Rep.~{\bf#1},\ #2 (#3)}
\def\PRD #1 #2 #3 {Phys.~Rev.~{\bf#1},\ #2 (#3)}
\def\PRL #1 #2 #3 {Phys.~Rev.~Lett.~{\bf#1},\ #2 (#3)}
\def\RMP #1 #2 #3 {Rev.~Mod.~Phys.~{\bf#1},\ #2 (#3)}
\def\ZP #1 #2 #3 {Z.~Phys.~{\bf#1},\ #2 (#3)}
\def\IJMP #1 #2 #3 {Int.~J.~Mod.~Phys.~{\bf#1},\ #2 (#3)}
\bibliographystyle{unsrt}

\newpage

\noindent
Fig.\ 1:  Distribution\cite{izen}
of difference between missing energy and momentum
for Cabibbo allowed and suppressed semileptonic decays.  The shaded regions
correspond to the background levels assuming (a-b) lead-proportional tube
calorimetry and (c-d) Cesium iodide crystal calorimetry.

\bigskip

\noindent
Fig.\ 2:  Missing mass distributions\cite{kim} for (a) $D^+\to\mu^+\nu$, (b)
$D_s^+\to\mu^+\nu$, (c) $D^+\to\tau^+\nu$ with $\tau\to\mu\bar\nu\nu$, and (d)
$D^+_s\to\tau^+\nu$ with $\tau\to e\bar\nu\nu$.  The shaded regions
correspond to the background levels.

\bigskip

\noindent
Fig.\ 3:  Reconstruction\cite{rafe} of the decay $D_s\to\eta\pi^+\pi^+\pi^-$
with $\eta\to\pi^+\pi^-\pi^0$ and $\pi^0\to\gamma\gamma$.

\bigskip

\noindent
Fig.\ 4:  Branching fraction for $D\to X + \ell^+\ell^-$ in the four generation
SM as a function of the CKM mixing factor, with the solid, dashed, dotted,
dash-dotted curve corresponding to
$m_{b'}=100, 200, 300, 400$ GeV, respectively.

\bigskip

\noindent
Fig.~5: \dm\ in (a) the four generation SM with the same labeling as in
Fig. 4, (b) in two-Higgs-doublet model
II as a function of $\tan\beta$ with, from top to bottom, the solid, dashed,
dotted, dash-dotted, solid curve representing $m_{H^\pm}=50, 100, 250, 500,
1000\gev$.  The solid horizontal line corresponds to the present experimental
limit.

\bigskip

\noindent
Fig.~6: (a) Constraints in the mass-mixing factor plane from \dm\
in the flavor changing Higgs model described in the text due to
the tree-level process (solid curve) and the box diagram (dashed).
(b) Constraints in the leptoquark coupling-mass plane from \dm.

\end{document}